# Report on power, thermal and reliability prediction for 3D Networks-on-Chip

Khanh N. Dang*[1], Akram Ben Ahmed[2], Abderazek Ben Abdallah[3], Xuan-Tu Tran[1]

**[1]**VNU-key Laboratory for Smart Integrated System (SISLAB), VNU University of Engineering and Technology, Vietnam National University, Hanoi, Vietnam
**[2]**National Institute of Advanced Industrial Science and Technology (AIST), Tsukuba, 305–8568, Japan
**[3]**University of Aizu, Aizu-Wakamatsu, Japan

**Abstract**

By combining Three Dimensional Integrated Circuits with the Network-on-Chip infrastructure to obtain 3D Networks-on-Chip (3D-NoCs), the new on-chip communication paradigm brings several advantages on lower power, smaller footprint and lower latency. However, thermal dissipation is one of the most critical challenges for 3D-ICs where the heat cannot easily transfer through several layers of silicon. Consequently, the high-temperature area also confronts the reliability threat as the Mean Time to Failure (MTTF) decreases exponentially with the operating temperature. Apparently, 3D-NoCs must tackle this fundamental problem in order to be widely used. Therefore, in this work, we investigate the thermal distribution and reliability prediction of 3D-NoCs. We first present a new method to help simulate the temperature (both steady and transient) using traffics value from realistic and synthetic benchmarks and the power consumption from standard VLSI design flow. Then, based on the proposed method, we further predict the relative reliability between different parts of the network. Experimental results show that the method has an extremely fast execution time in comparison to the acceleration lifetime test. Furthermore, we compare the thermal behavior and reliability between Monolithic design and TSV-based TSV. We also explorer the ability to implement the thermal via a mechanism to help reduce the operating temperature.

*Keywords:* Thermal dissipation, Reliability, 3D-ICs, 3D-NoCs.

## 1. Introduction

3D Networks-on-Chip (3D-NoCs), as a result of combining Networks-on-Chip (NoCs) [1] with 3D Integrated Circuit (3D-ICs) [2], is considered as one the most promising technologies for IC design [3]. By providing parallelism and scalability of the NoCs to 3D-ICs, we even obtain lower power consumption, shorter wire length while reduce the design area cost by several times. Among several 3D-ICs, Through-Silicon-Via which constitutes as inter-layer wire is one of the near-future technologies. Monolithic 3D ICs is another method to implement the 3D-ICs [30][34]. With both technologies, we expect to have multiple layers of the system. To support communication within the system, 3D-NoCs offer router-based infrastructure where the 3D mesh topology is used.

Despite several advantages, 3D-ICs and 3D-NoCs have to confront the thermal dissipation issue. The temperature variation between the two layers has been reported to reach up to 10°C in [4]. *Cuesta et al.* [5] also conduct an experiment of four-layer and 48 cores which gives the temperature variation up to 10°C between a single layer. The main reason of thermal dissipation difficulty in 3D-ICs is the top layers act like obstacles that prevent the heat could be dissipated by the heatsink. To solve this problem, fluid cooling [5] or thermal cooling TSV [6] have been proposed.

By having a higher operating temperature, it is apparent that 3D-NoCs easily encounter thermal throttling. Moreover, in terms of reliability, there is an expected acceleration in the failure rate (or reduction in Mean-time-to-Failure). For semiconductor devices, one of the most well know model of thermal impact in reliability is the Black's model [7] where the fault rate acceleration $\pi_T$ is:

$$\pi_T = A(J)^n \times e^{-\frac{E_a}{k_B T}} \qquad (1)$$

where $A$ is constant, $J$ is the energy, $k_B$ is Boltzmann constant, $E_a$ is activation energy and $T$ is the temperature in Kelvin. Here, we would like to note that the activation energy of Copper is much higher than CMOS material which makes TSV more vulnerable than the normal gate. Since TSV can act as a cooling device, TSV-based NoC has a lower operating temperature than Monolithic; however, TSV also has lower reliability. Therefore, the reliability differences between Monolithic and TSV-based 3D-ICs also need to be investigated.

While the thermal behavior could be extracted by performing the real-chip, reliability cannot be directly measured. Most industrial methods based on Black's model [7] in Equation 1 by baking the chip under high temperature to accelerate the failure [31][32][33].

In this work, we investigate the impact of the thermal dissipation difficulty of 3D-ICs by proposing a method to predict the temperature and MTTF of each region of the system. We first use the commercial EDA tools [8][9] to design and analyze the power and energy per data bit of 3D-NoC router. Then, we extract the number of bit and operating time of synthetic and PARSEC benchmarks to obtain the average power consumption of each router inside the network. We then use a thermal emulation tool named Hotspot 6.0 [10] to obtain the steady grid temperature of the system. By adopting the Black' model of reliability, the tool follows up with a reliability prediction of the system. By following the method, the designer can fast extract the potential hotspots inside the 3D-ICs and predict the potential of the vulnerable region due to high operating temperatures. The results also suggest the possible mapping of fluid cooling or thermal TSV insertion [5][21][22]. The contribution of this work is as follows:

- A platform to model the power, temperature, and reliability of any NoC systems. Here, we specify for 3D-NoC but the technique is general can be applied for the traditional planar NoC system.
- Analyze the reliability of Monolithic and TSV-based NoC. While TSV-based NoC has a lower operating temperature, TSV's material (Copper) has lower reliability.
- Exploration and comparison between different layout strategies and cooling methods.

The rest of this work is organized as follows. Section 2 first survey the existing works. Section 3 describes the proposed method in detail. Experimental results are discussed in Section 4. Finally, Section 5 concludes this work.

## 2. Related works

In this section, we summary the literatures related to our proposed method. We start with the power model and then present the work on thermal estimation. Finally, the reliability estimations for 3D-NoCs are presented.

### 2.1. Power modeling for 3D Network-on-Chip

To measure the power consumption of a 3D-IC, the straight forward method is to fabricate and set up a measuring system [15]. However, it is difficult to obtain such a system, especially designing and fabricate the chip is expensive and designers want to estimate the value before sending to production. Therefore, modeling the power consumption is a necessary step.

To model the power of any digital IC system, two major parts which are static and dynamic power is considered as follows:

$$P = P_{dynamic} + P_{static} = sf_cC_LV_{DD}^2 + I_{off}V_{DD} \qquad (2)$$

where $s$ is the switching probability (or activity ratio), $f_c$ is the clock frequency, $C_L$ is the load capacitance, $I_{off}$ is the leakage current and $V_{DD}$ is the supply voltage. Based on Equation 2, common EDA tools [8][9] can estimate the power consumption based on the parameter of the library and the switching activity. In fact, power estimation tool such as Primetime requires switching activity to obtain the most accurate result.

Using Equation 2 can estimate the power consumption of any circuit; however, for a fast prediction, the power consumption of NoCs can be obtained by its switching activity. By obtaining the number of flits went through the router during simulation, it can estimate the dynamic power consumption. Meanwhile, the static power consumption is constant for the same configuration (voltage, frequency, design). For instance, ORION 2.0 [16] model power consumption as dynamic and static power. Physical parameters such as wire length and leakage current are calculated to estimate the static power. In [17], the authors use regression to estimate the power consumption of the system based on the existing values. Other works in [18][19] also consider dynamic voltage frequency scaling in power consumption.

While these works can help estimate the power consumption of our system, we observe it is not the most accurate once because of the differences in design choice and library. Therefore, in this work, we propose our own power extraction method. We used the EDA tools to estimate the dynamic and static power and then combine with the switching of the routers in the used benchmarks.

## 2.2. Thermal behavior prediction for 3D Network-on-Chip

Once we obtain the power consumption of modules within a system, we can estimate the temperature the chip. HotSpot [20] is one of the first tool to help estimate the temperature grid. The $6^{th}$ version of HotSpot now can estimate the temperature of 3D-ICs. There are also different tools such as 3D-ICE [21] and MTA [22]. While MTA performs a similar task as Hotspot by using the finite element method, 3D-ICE focuses on the potential of liquid cooling. *Cuesta et al.* [5] also explored different layout strategies and liquid cooling for 3D-ICs.

## 2.3. Reliability prediction for 3D Network-on-Chip

By having the temperature of the system, we now can estimate the potential reliability. As we previously have shown, Black's model [7] in Equation 1 is one of the first models for CMOS. MIL-HDBK-217F of the US Military [12] also released its own model of reliability acceleration related to temperature. HRD4 from industry [13] and RAMP from academics [14] are the other two models to estimate the reliability of the system.

Among these model, HRD4 consider the reliability as the same for the chip bellow 70℃. The rest of the models follows the exponential acceleration with operation temperature (in Kelvin).

On the other hand, industrial approaches on reliability prediction [31][32][33] is to bake the chip to high temperature and measure the average time to failure of the samples. By using Black's model, they can estimate the potential lifetime reliability under normal temperature.

# 3. Proposed method

Figure 1 shows the proposed method for the thermal and reliability prediction of 3D-NoCs. We first built Verilog HDL of 3D-NoC. Then, synthesis and place and route are the following steps to obtain the layout, netlist file, wire length, and physical parameters.

We then perform post-layout simulation and use Synopsys Primetime to extract the power consumption of the system. Based on the number of data-bit, we further extract the energy per data bit. Then, we now can estimate the power consumption of all benchmarks by multiplying the obtained value with the number of bit per router per time. The power consumption of each router is taken to the temperature estimator tool (Hotspot 6.0) to obtain the temperature map. At the end of this step, we obtain all temperature maps of all benchmarks.

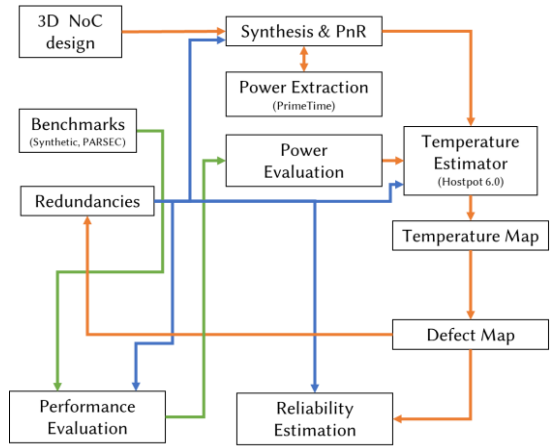

*Figure 1: Thermal and reliability prediction method of 3D Networks-on-Chip.*

One notable thing in 3D-NoCs is the possibility to have redundant Through-Silicon-Vias (TSVs). TSVs are usually made out of Copper and have a larger size than normal wire which can dissipate heat faster than normal silicon. Monolithic 3D-ICs fails to have the same feature since the via is extremely small. Consequently, we take the redundancy mapping into the hotspot prediction.

Once we can predict the temperature, we can obtain the reliability prediction using Black's model in Equation 1. Note that the activation energy also varies among materials. The output of reliability can also affect redundancy mapping as a close loop. Consequently, designers can further optimize the system to have the most balancing point of temperature, reliability, and area overhead. In the following part, we explained in detail each part of the proposed method.

We would like to note that our method reuses and follows the principle of existing works in academic and industrial approaches [8][9] [20][31][32][33].

### 3.1. Design of 3D Network-on-Chip

Here, we adopted our previous work in [3] with some modifications where the TSVs of a router is divided into four groups and place in four directions (west, east, north, south) of the router to support sharing and fault tolerance. However, we here provide more flexibility in the design since fault tolerance is not our objective of this work. Figure 4 shows the architecture of our 3x3x3 Network on Chip. Each router can connect to at most six neighboring routers in six directions and one local connection to its attached processing element. The inter-layer connections are TSV and we support optional redundant TSV group (yellow TSVs) which can be used to repair a faulty group group in the router. Borrowing and sharing mechanisms are another features we support to have high reliability in our system. More details on fault tolerance method can be seen in our previous work [3].

Each router receives a header flit of packet and support routing inside the network. Based on the destination, it forwards the header flit and the following flits (body and tail flits) to the desired port. Once the tail flit completes its transmission, the router starts to route the new packet.

In the router layout of [3], the design is not well optimized since it leases space between routers in layout. Figure 2(a) shows the layout of [3]. In order to optimize it, we use two different floorplans in this work. We first place TSVs and router logic in separated regions as in Figure 2 (b). Then, we place

TSV surround the router logics as in Figure 2 (c). We can notice that we reduce the size of the router significantly by removing the empty space.

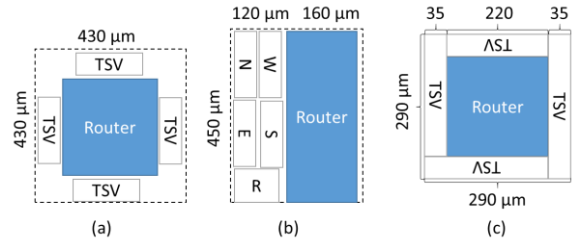

*Figure 2: Layout option for 3D-NoC router: (a) Previous work in [12]; (b) Separated TSV region; (c) Surround TSV region.*

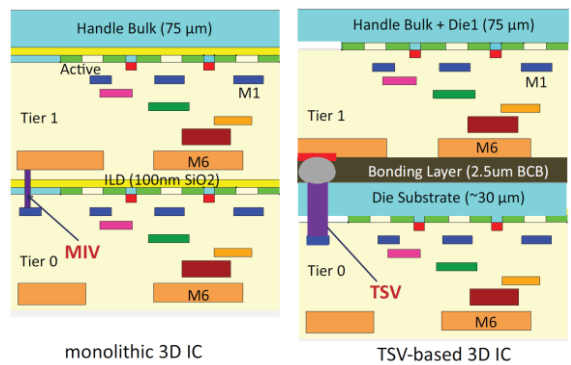

*Figure 3: 3D IC layer structure (heat sink on top) of Monolithic 3D IC vs TSV-based 3D IC [35].*

Among the two new layouts, Figure 2(c) provides the best thermal balance because it isolates the logic of a router to the nearby module. Since routers are usually hotspots inside the system, placing them near a hot area can raise it temperature significantly. Here, by surrounding by TSVs, we create isolation for the router. Furthermore, Copper has low thermal resistivity which can dissipate the heat from the router to the upper layers. By doing so, we can transfer then heat to the top layer and the heatsink. In the evaluation section, we then discuss the efficiency and cost of inserting thermal via in our design.

Figure 3 shows the different between Monolithic and TSV-based 3D-ICs. While TSV is made out of Copper that dissipate thermal faster than Silicon layers. However, there are bonding layers between stacking using TSV which creates an isolation of thermal disspation between them.

Our 3D-NoC also support fault tolerance for not only TSVs but also for soft and hard fault [36-47]. For soft error in data, we used two SECDED(22,16) to detect and correct two and four faults [36][38], respectively. For pipeline stage, we support double execution for detection and correction using a third execution [36][45][46]. We also tolerate permanent faults in crossbar, buffer and link [36][38][48]

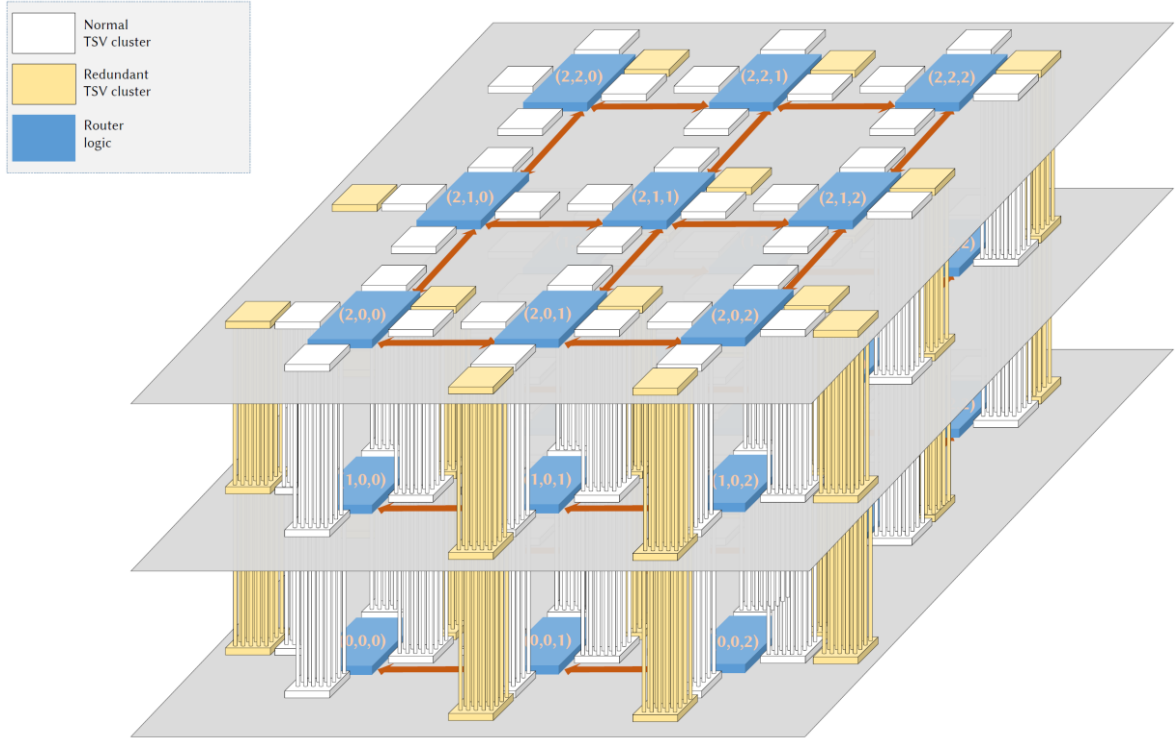

*Figure 4: Architecture of our 3D Network-on-Chip with the size of 3x3x3.*

### 3.2. EDA tools and power extraction

The following part of the method is to use EDA tool to extract the power consumption. Apparently, we can use any supported EDA to obtain power consumption. For our experiment, we use Synopsys Design Compiler, ICC and Primetime to do the physical design and extract the power consumption.

To extract the power, we perform a heuristic transmission benchmark of a single router. Here, we generate two packets of ten flits in all possible directions. Because our router supports returns the flit from it sending ports, we have 7x7=49 possible directions. By using PrimeTime, we can obtain the dynamic and static power.

Here, we also classify the energy into static and dynamic. While static power consumption is stable, we keep the value as it is. For the dynamic power, we calculate the total energy and the energy per data bit.

### 3.3. Power and Temperature Estimation

Once we obtain the energy per data-bit, we can obtain the overall power consumption as follows:

$$P = P_{static} + P_{dynamic} = P_{static} + E_{dynamic} \times \frac{N_{bit}}{time} \tag{3}$$

where $N_{bit}$ is the number of a data bit in the benchmark. We can also scale the power with the dynamic frequency and voltage if needed. Here, we also support dynamic scaling for voltage and frequency by using Equation 2 where different voltage and frequency can be converted using the following equations:

$$\frac{P_{static}^{V_1 F_1}}{P_{static}^{V_2 F_2}} \approx \frac{V_1}{V_2} \tag{4}$$

$$\frac{P_{dynamic}^{V_1 F_1}}{P_{dynamic}^{V_2 F_2}} \approx \frac{f_1 V_1^2}{f_2 V_2^2} \tag{5}$$

where $V_1, f_1$ and $V_2, f_2$ are two pairs of supply voltage and frequency.

The power trace and floorplan are taken into Hotspot 6.0 to obtain the thermal map of the design. The results of Hotspot 6.0 is the steady temperature of each router and its TSVs. We can also support transient power and temperature. However, since we consider the reliability as the major target, the steady temperature is the most important value.

### 3.4. Defect mapping

After getting the thermal map, we can extract the reliability to obtain the defect map. Figure 6 shows the normalized thermal acceleration model in academics and industry. We illustrate the MIL-HDBK-217F of the US Military [12], HRD4 from industry [13] and RAMP from academics [13]. Notably, we used Black's model [7] in our work. However, we could also adopt the existing model if needed as in Figure 6. One common between the model is the exponential curve of acceleration of the fault rate with the temperature. Note that HRD4 uses 70°C as the threshold of reliability concern.

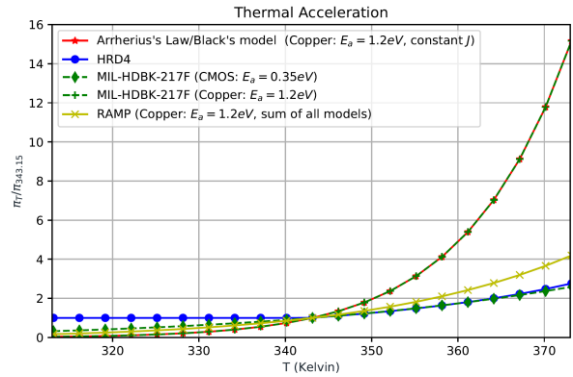

Figure 6: Normalized thermal acceleration of fault rate.

Table 1: Normalize fault rate of Copper TSV mapping using Black's model [7].

| Temperature (K) | Normalize fault rate to 70°C |
|---|---|
| 303.15 | 0.011537 |
| 313.15 | 0.039174 |
| 323.15 | 0.123317 |
| 333.15 | 0.362371 |
| 343.15 | 1 |
| 353.15 | 2.605435 |
| 363.15 | 6.439561 |
| 373.15 | 13.94691 |

Table 1 shows the fault rate mapping obtained by Black's model [7]. At 30°C, the fault rate is less than 2% at 70°C. However, once the IC operates at 80°C, its fault rate is 2.6× at 70°C and 220× at 30°C. By mapping to fault rates, we can find the critical part of the 3D-NoCs in terms of reliability.

## 4. Experimental results

In this section, we evaluate the 3D Network on Chip [3] using the proposed platform. Furthermore, explore the idea of the different floorplan and cooling strategies. At first, we extract the power consumption from the synthetic benchmark of a router. Then, we estimate the power consumption of the 3D-NoC system under various benchmarks. Then, temperature and reliability prediction is illustrated. In the final part, we compare different strategies for layout and cooling.

### 4.1. 3D-NoC router power estimation.

We used the previous work [3] model of a router to estimate the power consumption and the energy. Note that we modified the router with some optimizations and further fault tolerances. We use NANGATE 45nm library [23] and NCSU FreePDK TSV [24]. The hardware complexity of the router is shown in Table 2. We perform a heuristic benchmark for this router by sending each port to all possible ports two packets of ten flits of 32 bits. The number of bits is $7 \times 7 \times 2 \times 10 \times 32 = 31360$ bits. The desired injection rate is 1 flit/port/cycle. The final result for static power and energy per data bit are 7.66e-4 W and 9.246e-13 J/bit, respectively.

*Table 2: Hardware complexity of our 3D-NoC router.*

| Parameter | Value |
|---|---|
| Area cost | 38,838 $\mu m^2$ |
| Maximum Frequency | 537.63 MHz |
| Operating Frequency | 500MHz |
| Technology | 45nm (NANGATE 45) [23] |
| Voltage | 1.1 V |
| Static Power (at 500MHz) | 7.64e-4 Watt |
| Dynamic Power (at 500MHz) | 1.028e-2 Watt |
| Simulation time | 2.823200e-6 second |
| Energy | 2.9022496e-8 Joule |
| Energy per data bit | 9.2546e-13 Joule/bit |

### 4.2. 3D-NoC system power estimation.

To estimate the power of 3D-NoC system, we use Equation 3 with the scaling Equation 4 and 5 for different voltage and frequency if needed. Apparently, we need to obtain the number of the bit through the routing during its operation. Here, we perform both synthetic benchmarks (Matrix, HotSpot, Uniform, and Transpose) from [3] and we design a 3D-NoC version of garnet 2.0 in gem5 [25] then perform the PARSEC benchmarks suite [26]. PARSEC is one of the most well-known benchmarks for multi-core computing systems. Here, we use 64 core x64 processors as the processing elements of the PARSEC benchmarks. Here, we only extract the number of flits that went through the routers to estimate the power consumption. The power consumption of the processing elements can be obtained by using McPAT tools by HP inc. [27]; however, it is out-of-scope of this work.

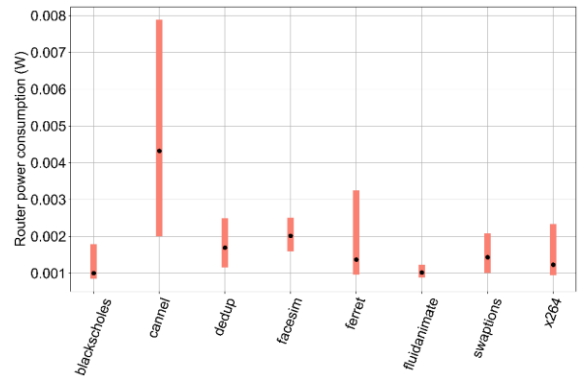

*Figure 7: Power consumption of our 3D-NoC under PARSEC benchmarks.*

Figure 7 shows the power consumption of our 3D-NoC under PARSEC benchmark. Here, we scale the frequency to 2GHz to fit with the configuration of gem5 using Equation 4 and 5. Among these benchmarks, we observe cannel has the highest power consumption and also the highest variation (between the minimum and maximum power of router).

Figure 8 shows the power consumption of the 3D-NoC system under synthetic benchmarks. We keep the frequency as 500MHz and inject the flit with a maximum inject rate. Note that we perform two Hotspot benchmarks where two nodes are the destination of 5% and 10% of total flits. We can easily observe the significant drop when increasing the number of flits to the hotspot nodes. This can be explained by the congestion created by more flits coming to these nodes which extend the execution time of the system. On the other hand, the matrix benchmark has the lowest router power consumption. We also notice that the synthetic benchmarks have much higher power consumption than the PARSEC benchmarks since no computation is taken in this benchmark. As a consequence, the execution time is shorter which makes the power consumption higher than PARSEC.

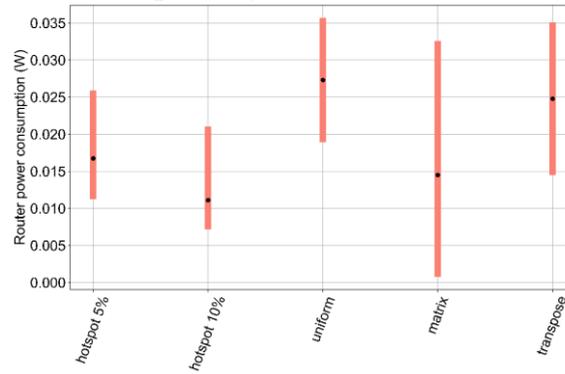

*Figure 8: Power consumption of our 3D-NoC under synthetic benchmarks.*

### 4.2. 3D-NoC thermal estimation.

By using the power estimation of the previous section, we conduction the thermal estimation using Hotspot 6.0 [10]. Table 3 shows the configurations for thermal estimation using Hotspot 6.0. We modify the thermal resistivity corresponding to our designed TSV (Copper with the size of $4.06\mu m \times 4.06\mu m$) using the following equation [28]:

$$R_{joint} = \frac{Area}{\dfrac{Area - Area_{TSV}}{R_{TIM}} - \dfrac{Area_{TSV}}{R_{Copper}}} \qquad (6)$$

where TIM is the thermal interface material. The result of the thermal resistivity of the layout in Figure 2(c) can be found in Table 3. The final TSV area thermal resistivity is 0.0226mK/W.

*Table 3: Configurations for thermal estimation.*

| Parameter | Value |
|---|---|
| Router floor-plan | 290 $\mu m$ ×290 $\mu m$ |
| Floorplan | Figure 2(c) |
| One TSV area | 4.06μm×4.06μm |
| Router logic area | 220 $\mu m$ ×220 $\mu m$ |
| Router logic utilization | 80% |
| TSV area/utilization | 35,700 $\mu m^2$ / 10.16% |
| Copper thermal resistivity | 0.0025mK/W |
| TIM thermal resistivity | 0.25mK/W |
| TSV area thermal resistivity | 0.0226mK/W |

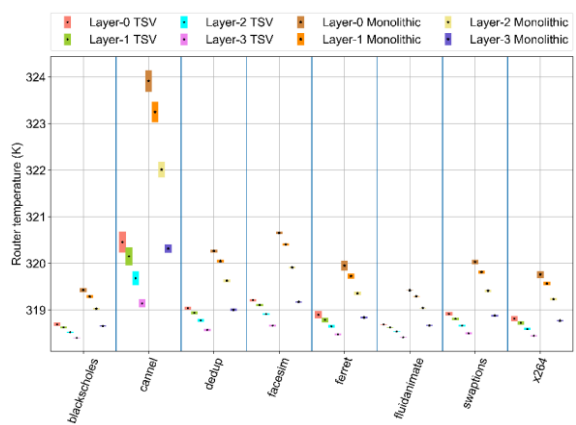

*Figure 9: Temperature of our 3D-NoC under PARSEC benchmarks.*

To compare with Monolithic 3D-IC, we also adopt the method in [35] where we remove the bonding layers between silicon layers. We keep the thickness of the silicon layer as it is for a fair comparison. Obviously, if we thin the layer, the transfer of heat is much faster.

Figure 9 shows the router temperature under PARSEC benchmark. Here, we also compare with the monolithic technology where no TSV needed [33]. As we can observe in Figure 9, TSV-based system has lower operating temperature thanks to the ability to transfer the heat of Copper TSVs. The difference in temperature is around 1K at the bottom layer and even reach 3.5K in the cannel benchmark.

Figure 10 shows the operating temperature under synthetic benchmarks of our 3D-NoC. We can easily notice that the operating temperature of Monolithic systems is much higher than TSV ones since we stress the system under it saturation points. The highest temperature of Monolithic 3D-NoC even reaches 351.64 K (78.49°C). The hottest layer of the TSV-based system has a similar temperature as the coolest layer of Monolithic 3D-NoC.

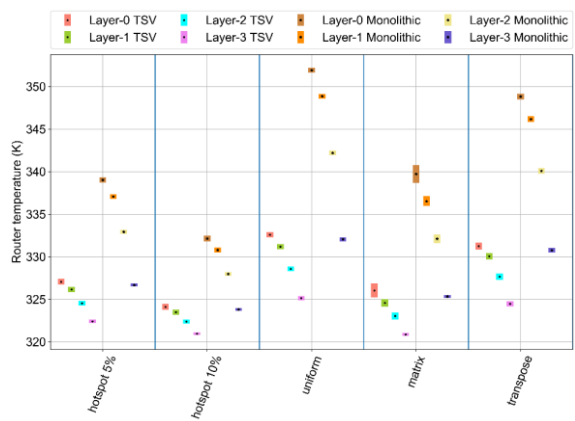

*Figure 10: Temperature of our 3D-NoC under synthetic benchmarks.*

### 4.2. 3D-NoC reliability estimation.

In this section, we use Black's model to evaluate the MTTF of 3D-NoC. Figure 11 and Figure 12 show the normalized MTTF of each layer to 323.15K (50°C) under PARSEC and synthetic benchmarks. Here, we can observe TSV-based 3D-NoC dominates Monolithic in the PARSEC benchmark. With synthetic benchmarks, TSV-based 3D-NoC is slightly better than Monolithic ones.

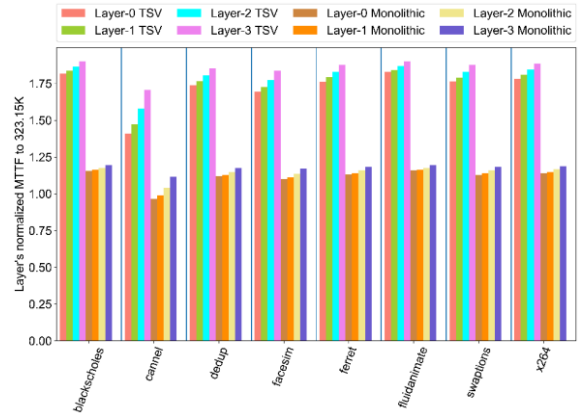

*Figure 11: Normalized MTTF of our 3D-NoC under PARSEC benchmarks.*

## 4.4. Exploring different layout and thermal dissipation method

In this section, we explore different layout and thermal dissipation for our 3D-NoC. First, we perform thermal and reliability prediction for our layout in Figure 2(b). Then, we insert four thermal TSVs with the size 15 $\mu m$ ×15 $\mu m$ in four corners of the router floorplan in Figure 2(c). This size of TSV is still feasible in the existing manufacture process[34]. We also add 10 $\mu m$ Keep-out-Zone distance this thermal TSV to avoid mechanical stress. The thermal TSV went through all layer of TSV but does not contact with the heatsink. The heatsink and thermal TSV are separated by a layer of thermal interface material.

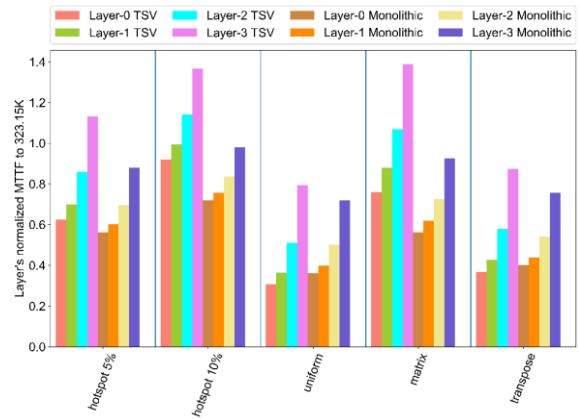

*Figure 12: Normalized MTTF of our 3D-NoC under synthetic benchmarks.*

Figure 13 and Figure 14 show the thermal behaviors under PARSEC and synthetic benchmark for different layouts and cooling. We can notice that with the layout in Figure 2(b) has the worst thermal behavior among the TSV designs. On the other hand, adding thermal TSV can help reduce the operating temperature significantly. By adding four TSVs, we can even reduce the temperature by nearly 1K at the bottom layer in the uniform benchmark which is the most stressed benchmark. Another benchmarks' results also show a slight improvement in thermal behaviors.

One thing we can easily notice the top layer's temperatures do not change. This is due to the fact it is already cool down by the heatsink and adding TSV cannot help it reduce the temperature. Also, the heatsink temperature is raised near the top layer temperature which reduces the ability to transfer heat. If the thermal TSV can contact the heatsink, it can significantly cool down the bottom layer. Also, liquid cooling could be extremely helpful in this situation.

In comparison to the traditional 2D-ICs, we observe that the TSV-based ICs have higher operating temperatures. The 2D-based 3D-NoCs operate under 319K and 322K with PARSEC and synthetic benchmarks, respectively. On the other hand, TSV-based system increases at most 10K in maximum temperature with the layout in Figure 2(b).

In summary, different layouts can make different thermal behaviors. The layout Figure 2(b) does not surround the router by TSV area, therefore, the router could heat up each other and reach a higher temperature. On the other hand, adding thermal TSV to cool down the bottom layer is helpful since it can reduce nearly 1 Kelvin in the worst case. By mapping to the reliability, we can easily obtain a 2×~3× improvement of MTTF.

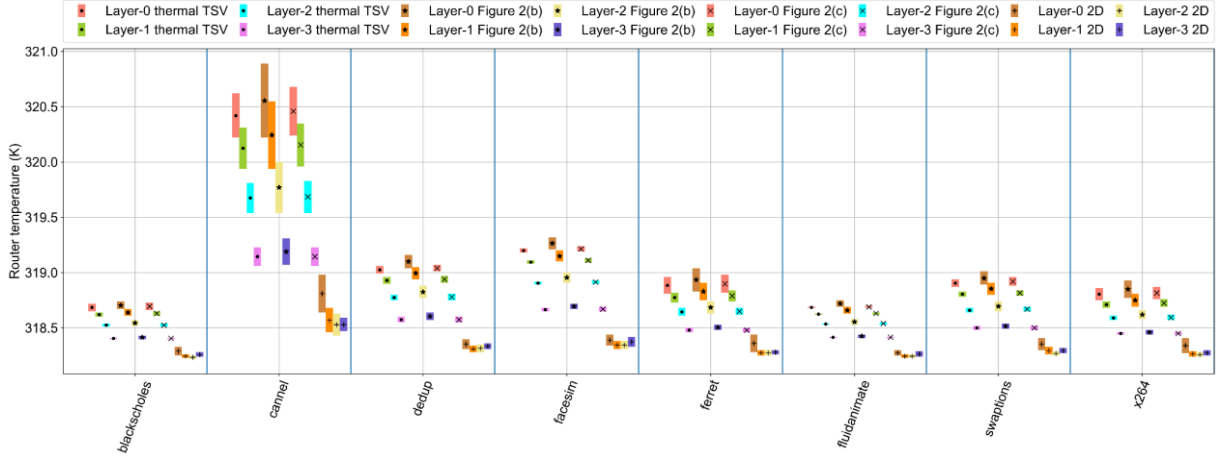

*Figure 13: Thermal behavior of different layout and cooling methods under PARSEC benchmark.*

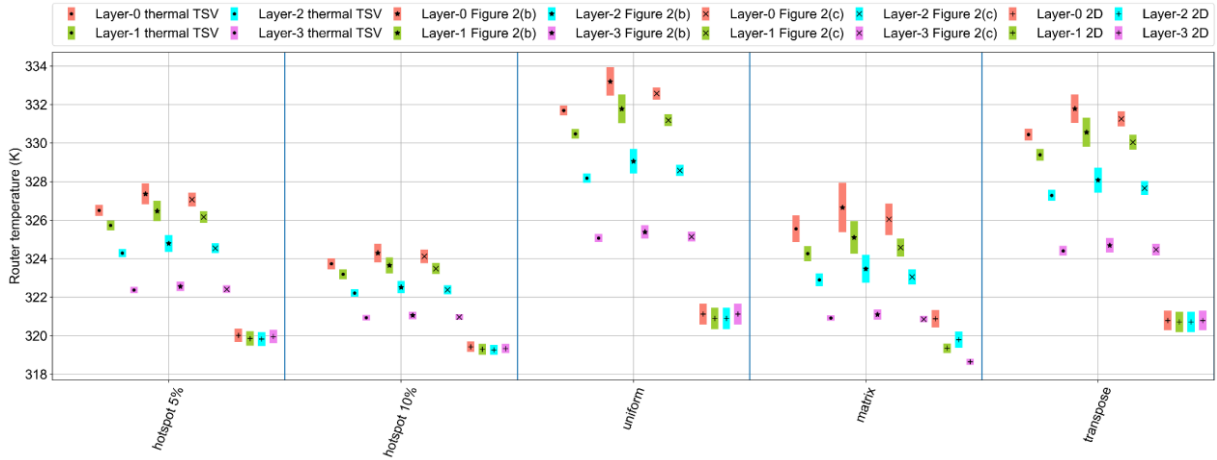

*Figure 14: Thermal behavior of different layout and cooling methods under synthetic benchmarks.*

## 4.5 Execution time

In this work, we evaluate the proposed method using a system with Xeon E5-2620 8 cores 2.1GHz, 16GB RAM and Linux Subsystem and PowerShell under Windows 10. The platform is written under C++, Python, and Bash. The execution time is measured using command time under Linux and Measure-Command under Windows PowerShell. Here, the simulation time of PARSEC and synthetic benchmarks are not considered because they are separated from our own flow. As shown in Table 4, all steps in our flow perform under two seconds. Our method easily outperforms in terms of execution time the fabrication-based methods which usually take hours regardless of designing, fabrication and assembly time [31][32][33].

*Table 4: Execution time of the proposed flow.*

| Work | Step | Time |
|------|------|------|
| Ours | Power extraction (one benchmark) | 1.22 s |
| | Floorplan generate | 0.095 s |
| | Temperature estimation (one benchmark) | 81 s |
| | Reliability estimation (12 benchmarks) | 1.12 s |
| [31] | Reliability test | 96h |
| [32] | The longest step in reliability test | 1000h |
| [33] | Lifetime acceleration test | 100-5000h |

### 4.6. Discussion

In this section, we would like to discuss some technical details of our methods. Advantages and drawbacks are also mentioned in this part.

In our evaluation, we point out that Monolithic has a higher temperature than TSV-based 3D-NoC due to two major reasons: (1) TSVs act like thermal conduct devices and (2) Monolithic 3D-ICs has a higher density than TSV-based system. However, we would like to note that Monolithic 3D-ICs has lower area cost than TSV-based system.

Fluid cooling [5] is one of the most advanced methods to reduce the operating temperature of the system. Although we have not explored the ability of this method, it has shown promising efficiency for 3D-ICs[5]. With a fast velocity of the fluid, we expect the system can be cooled down significantly. However, we would like to note that fluid cooling has unknown reliability which needs to be carefully investigated for being widely used.

## 5. Conclusion

In this work, we proposed a platform to quickly estimate the power, thermal behavior, and reliability of 3D-NoC systems. The method has shown extremely quick execution time. We also analyze and simulate the reliability of TSV and Monolithic 3D-ICs. Furthermore, we explore and compare different layout strategies and cooling methods.

In the future, advanced cooling such as liquid could be investigated. The impact of DVFS and fault tolerance on performance and thermal behavior also could be studied.

### Acknowledgment

This research is funded by the Vietnam National Foundation for Science and Technology Development (NAFOSTED) under grant number 102.01-2018.312.